# Crowdsourced Peer Learning Activity for Internet of Things Education: A Case Study

Ahmed Hussein, *Cardiff University, United Kingdom*
Mahmoud Barhamgi, *Claude Bernard University Lyon-1, France (mahmoud.barhamgi@univ-lyon1.fr)*
Massimo Vecchio, *FBK CREATE-NET, Italy (mvecchio@fbk.eu)*
Charith Perera *Cardiff University, United Kingdom (charith.perera@ieee.org)*

**ABSTRACT**

Computing devices such as laptops, tablets and mobile phones have become part of our daily lives. End users increasingly know more and more information about these devices. Further, more technically savvy end users know how such devices are being built and know how to choose one over the others. However, we cannot say the same about the Internet of Things (IoT) products. Due to its infancy nature of the marketplace, end users have very little idea about IoT products. To address this issue, we developed a method, a crowdsourced peer learning activity, supported by an online platform (OLYMPUS) to enable a group of learners to learn IoT products space better. We conducted two different user studies to validate that our tool enables better IoT education. Our method guide learners to think more deeply about IoT products and their design decisions. The learning platform we developed is open source and available for the community.

**INTRODUCTION**

Most of the technically savvy (even some of the non-technical) end users know primary components of a computer (or a table or a mobile). For example, some of the major consideration factors towards buying these devices could be CPU, memory, battery, screen size, brand, etc. Traditional computing devices such as above are general purpose devices. Their functionalities and characteristics are well known, which makes it easier for comparison and understand. However, we cannot say the same for the Internet of Things (IoT) devices [1] [2], due to lack of standardisation, lack of mature market segmentation and heterogeneity.

IoT has many different market segments such as smart home, smart wearable, and so on. Further, each of these segments can be divided into different subsegments such as smart lighting and smart activity monitor, and so on. Furthermore, even within smart activity monitor, there could be devices that primarily focus on fitness and others may focus on health. As a result, there are a plethora of IoT devices in the market. Due to heterogeneity, there isn't an easy way to compare these products from design decisions point of view. The question (or the gap) we wanted to address was *'How can we develop a systematic method to analyse and examine IoT products in the market with the intention of learning and critically analysing their design decisions'*.

Our solution to the above problem was to design and develop a crowdsourcing-based platform that facilitates peer learning activities. Our approach allows users (we call them 'learners') to collectively undergo an entire learning experience where they will learn about different IoT products, design decisions, major characteristics, and IoT in general. Our proposed method was inspired by three areas, namely, crowdsourcing [3], learning technology, and IoT.

**CROWDSOURCING**

The definition of crowdsourcing can come in several forms, with debates surrounding its criteria. Jeff Howe [4] claims it is the combination of two words, "crowd" and "outsourcing". While crowdsourcing is much cheaper than outsourcing, it allows organisations or simply people, to tap into the talent of a large crowd. An example of crowdsourcing used in education domain is CrowdyQ [5], a service for creating well-written examination papers. Users collaborate to develop high-quality exam questions. This platform addresses the issue of the unavailability of skilful exam developers.

According to Bar and Maheswaran [6], crowdsourcing can be split into two genres, explicit and implicit. In explicit systems, collaborators build information artefacts. Alternatively, implicit systems enable users to solve problems for its owner. In addition to this argument, the level of effort required to maintain a system could also be used to identify a crowdsourcing platform. An example of an explicit system is Linux or Wikipedia. Wikipedia is a huge encyclopaedia authored by crowdsourcers. Daren C. Brabham [7] lists some key identifiers of a crowdsourcing platform: 1) an organization that has a task it needs to be performed, 2) a community (crowd) that is willing to perform the task voluntarily, 3) an online environment that allows the work to take place and the community to interact with the organization, and 4) mutual benefit for the organization and the community.

Estelles-Arolas et al. [8] has defined crowdsourcing as *"a type of participative online activity in which an individual, an institution, a non-profit organization, or company proposes to a group concepts, theories, and cases to individuals of varying knowledge, heterogeneity, and number, via a flexible open call, the voluntary undertaking of a task"*. Crowd worker typically receives satisfaction through economic means, social recognition, self-esteem, or the development of individual skills.

**LEARNING TECHNIQUES FOR INTERNET OF THINGS**

Many teaching and learning techniques have been developed to engage students of all abilities. For instance, splitting up larger problems or assignments into smaller, more approachable tasks makes the initial job far less daunting. This is common within computing, often likened to problem-solving. A meta-analysis by Dunlosky et al. [9] have evaluated ten of the most popular teaching techniques, then categorised them in terms of utility in absorbing learning material. For instance, "*highlighting, marking, underlining, summarising and re-reading*", while popular forms of studying, ranked as low utility. Methods that were ranked as moderate utility were, '*Elaborative Interrogation* (the process of asking yourself why in an attempt to understand concepts), '*Self-Explanation*' (involves the participant explaining and recording how one reaches an answer or conclusion) and '*Interleaved practice*' (is when the student studies the topic at hand but also blends the study with previous topics/concepts at the same time).

Our proposed technique was designed with these techniques in mind. Ideation cards and card games are increasingly used in education domain from kindergarten to universities to professional development. More importantly, as presented in Figure 1, card-based learning activities are increasingly using to teach IoT. Our method is greatly influenced by card games and underlying principles of such activities.

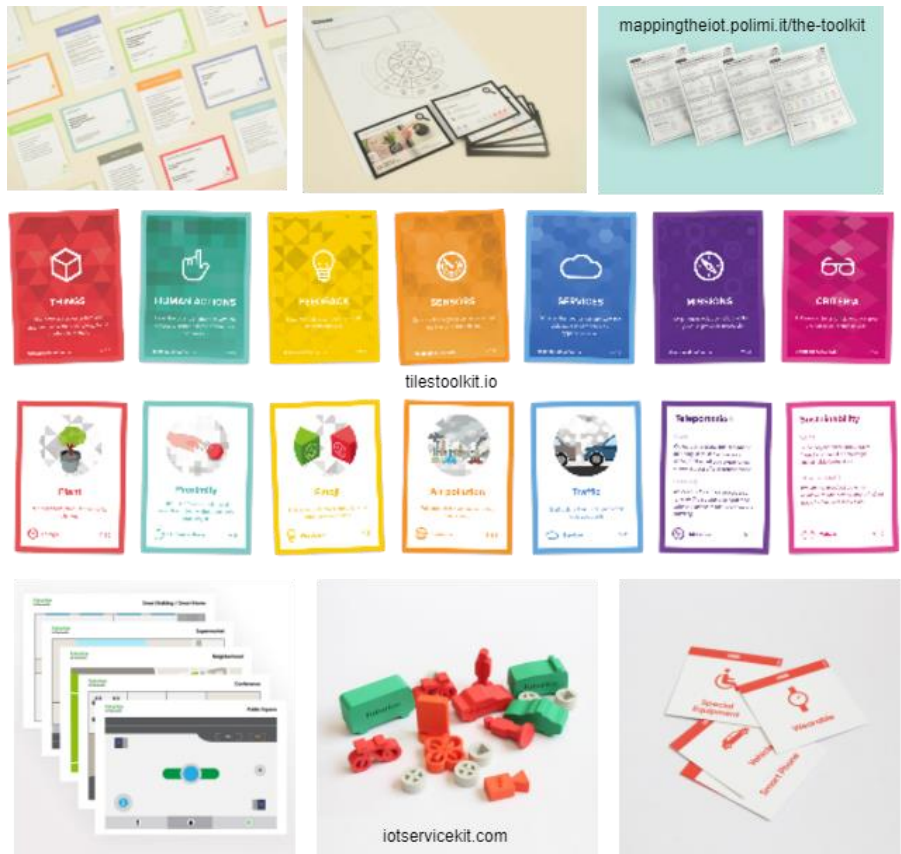

*Figure 1: Card based activities developed to support the Internet of Things Education*

**PROPOSED APPROACH: CROWDSOURCED PEER LEARNING ACTIVITY**

The proposed learning method consists of three stages. The first stage, also known as the **(1) Crowdsourced Investigation Stage**, involves learners analysing and identifying the technologies that a given IoT product is comprised of. To ensure the IoT products are fully analysed, each device is analysed by multiple learners (independently). For example, as presented in Figure 2, each user may investigate different sources such as packaging, web pages, promotional videos, advertisements, app user interfaces, terms and conditions, leaflets and so on. The critical step is to collect evidence. It is useful that learner examines different sources (e.g., finds out a given IoT product have an accelerometer sensor embedded into it). However, what is more important is to collect evidence about a given fact. For example, what we want learners to do is not only to find out important aspects/characteristics of a given IoT product but also gather much evidence possible. The objective of this stage is to create IoT product profiles that consolidate information (i.e., design decisions, major characteristics, features) backed by evidence. In the next section, we explain how we implemented this concept into the proposed learning platform. We believe that such an investigation helps learners to understand the characteristics of IoT products and their design decision better. Such familiarity is critical in the next two stages.

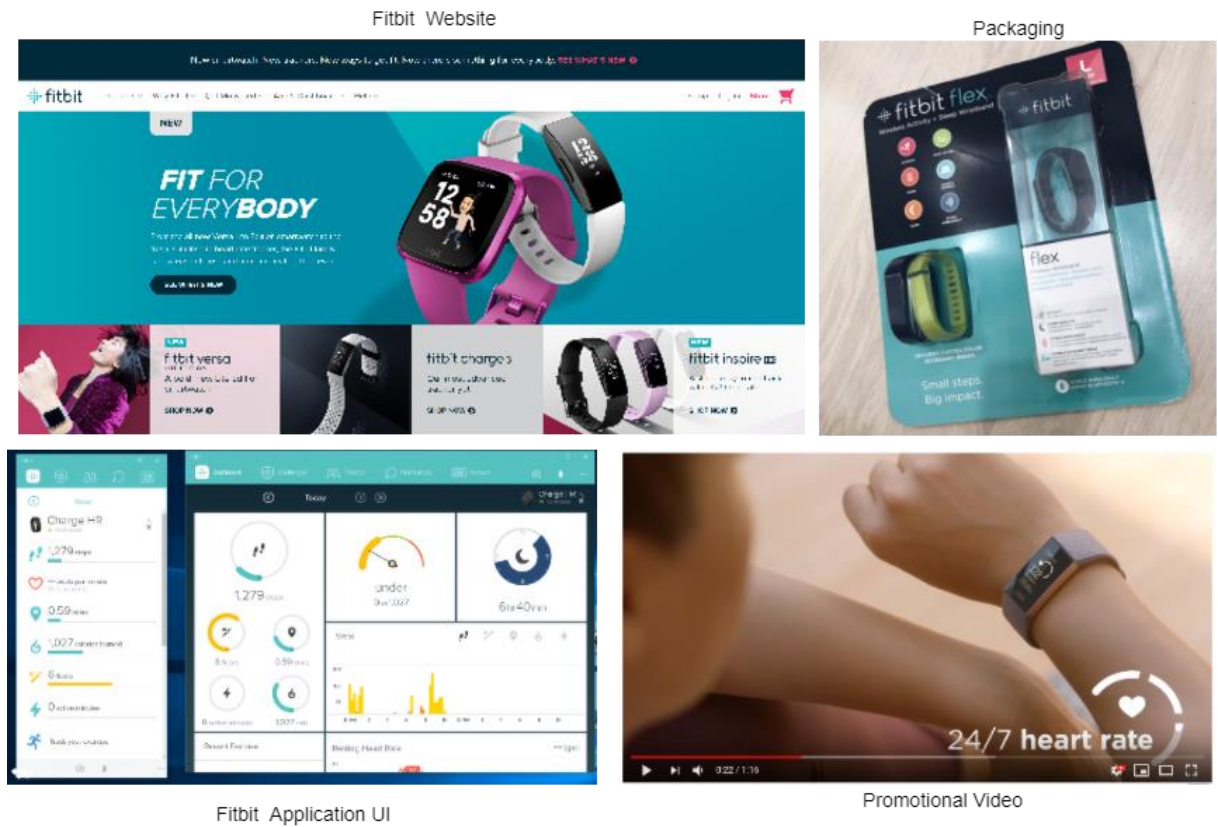

*Figure 2: Different source provides different information about how IoT devices are being designed and built (e.g., what sensors are integrated in to the product, what kind of communication technologies are used, etc.)*

The next stage is **(2) Merging Stage**. In stage one, each learner independently creates IoT product profiles by identifying characteristic and collecting evidence to support their claims. To ensure the devices have been fully analysed, each device is analysed by multiple learners. As a result, at this point, multiple learners may have created multiple profiles of the same IoT product independently. To facilitate the above approach, we propose our learning activity be performed by a few different learners at the same time. This is important in order to formulate a peer learning environment. At the merging stage, learners are expected to gather around as a team where in order to collectively perform the merging process.

In this stage, learners need to discuss and merge the best possible choices into a product profile. For example, let's say three learners have created three different product profiles analysing Fitbit [2] devices. Two of the learners may have found that Fitbit is using Bluetooth as communication technology. However, the third learner may have found out that Fitbit also uses Wi-Fi communication to update its firmware (i.e., two learners have missed this information). At this point, the three learners need to discuss their findings. Ideally, they may decide to add both Wi-Fi and Bluetooth as supported communication technologies into their merged IoT product profile (for Fitbit).

Similarly, learners, as a group, need to go through each feature/characteristics/design decision in order to create a consolidate master profile for each IoT product. We believe such discussion could lead to independent learning, which we also identified as deep leaning [10]. Deep learning happens through activities such as *'learning by example', 'learning by doing', 'learning by reviewing' and 'learning by discussing your experiences with others'*.

The final stage, the **(3) Comparison Stage**, is where compare different types of IoT products. In this stage, the discussion is around feature, characterises across different IoT products. For example, Fitbit may have used a heart rate sensor. Similarly, Beddit [1], [2] device may also have embedded a heart rate sensor. The conversation we would like the learners to have is around design decision. For example, some sample question we want them to discuss are: (i) why there is an accelerometer sensor is integrated into a given product? (ii) what kind of information a given product may extract using this particular type of sensor? (iii) why the battery of a given IoT device can last long and why some devices are draining the battery quite quickly? (iv) why a particular IoT device is using a particular type of communication protocol, and so on. Questions are pretty much unlimited. What we would like the learner to get out of this phase is to understand how IoT devices are built, what kind of decisions product designer and developer might need to make, and critically review those design decisions.

We acknowledge that these discussions can never be perfect, due to many different factors such as lack of availability of information and lack of knowledge of learners. However, our objective is not to give perfect knowledge to the learner. Most of the time it is difficult to find out how a given IoT product manages data (e.g., what kind of data it collects, how frequently, where data is being stored, how long they are stored, how such data will be transformed to knowledge and insights throughout the data pipeline). Therefore, our primary objective is to guide and encourage learners to analyse and compare IoT products critically and try to understand the differences towards building an overall understanding on technical and social challenges that products face or may face in the future. In the next section, we will discuss how we implement these stages in the proposed platform.

To summarise, our proposed crowdsourced peer learning activity has three stages as depicted in Figure 3, namely, crowdsourced investigation stage, merging stage, comparison stage.

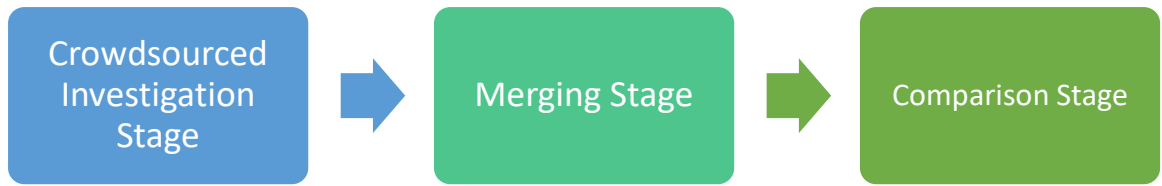

Figure 3: Stages in the proposed crowdsourced peer learning activity for IoT

**OLYMPUS: CROWDSOURCING PLATFORM FOR PEER LEARNING INTERNET OF THINGS**

The three separate stages we discussed earlier are implemented around three different user login roles, namely, (i) crowd worker (crowdsourced investigation stage), (ii) admin (merging stage), (iii) student (comparison stage). Learners are expected to play all three different roles as explained below. The learning activity begins with admin create a master template for a selected IoT product where the learners need to fill in. For example, the admin may create an IoT device template and add basic information such as product name, description, and brand. This template is then available to all crowd workers. Crowd workers are then guided through a step by step process. Our platform allows crowd workers to enter information about features of their choice complemented by evidence. Crowd workers may choose built-in features and prebuild choices (e.g., connectivity as a feature and Bluetooth 4.0 as the answer). Crowd workers then can add evidence. For example, the learner may upload a PDF and mark a particular page or may upload a video and mark a particular time frame). Crowd workers can also create new features of their choice if the pre-built features are not sufficient to capture the interesting aspects of a given IoT product. Figure 4 depicts how a Crowd worker might add a feature and relevant evidence to an IoT product profile.

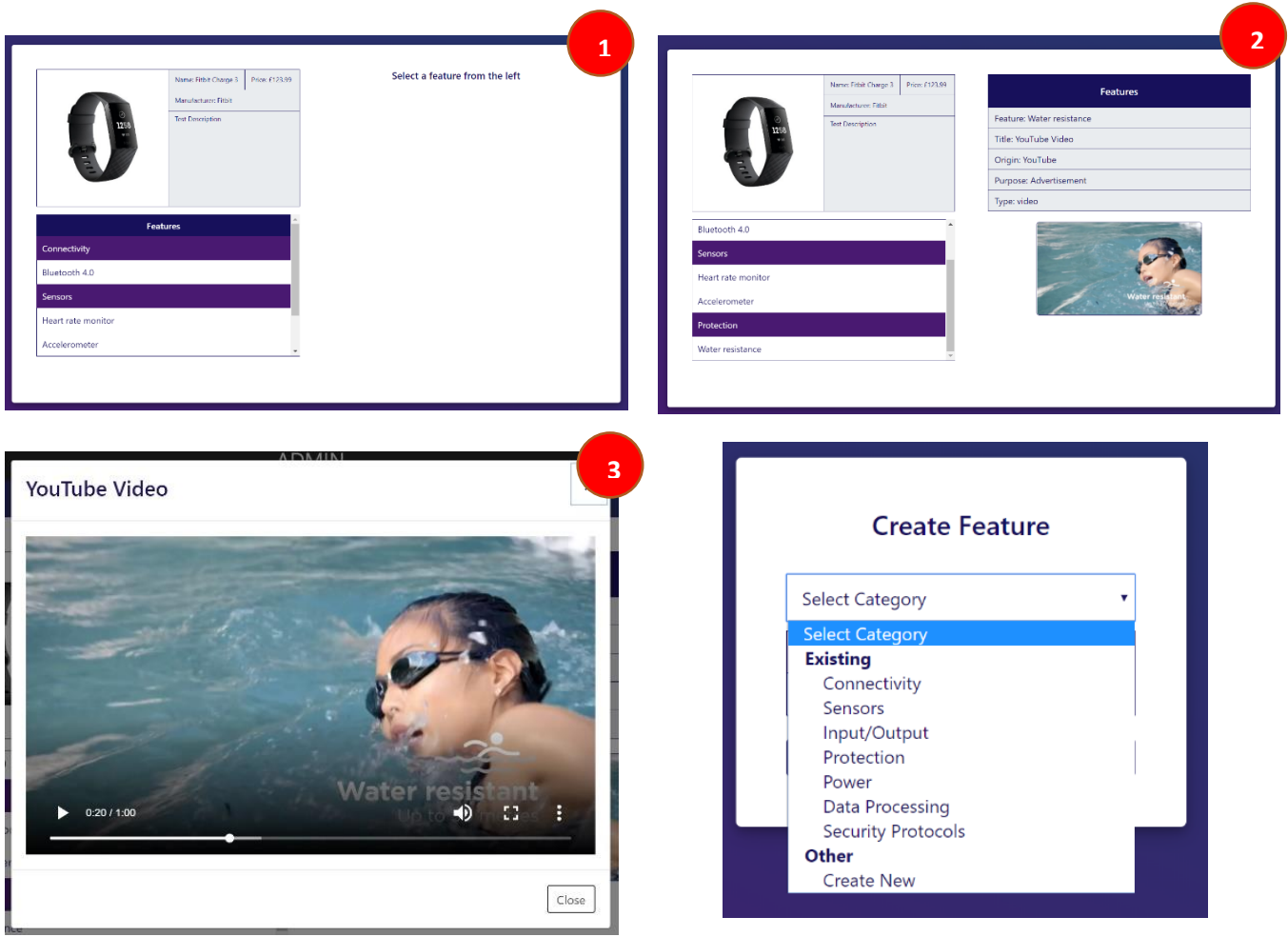

*Figure 4: These series of screens demonstrate how a crowdsource worker add feature of water resistant to given IOT device and provide video-based evidence*

When the learners are logged in as admins, they are presented with a step-by-step workflow (similar to stage 1). In this case, learners are asked to pick the version they want to merge into the master IoT product profile. For example, in Figure 5, learners are presented with alternative sources to pick as relevant evidence. It is important to highlight that non-competing features are always pre-merged (with the option to remove). This means that, if one crowed worker provides evidence related to a heart rate monitor and another worker refers to an accelerometer, both features will be merged by default. However, if more than one crowd worker provides evidence related to a common feature, then the learners are asked to pick the best evidence. We believe this process encourages learners (as groups) to discuss each feature and its impact on the given IoT product, in details.

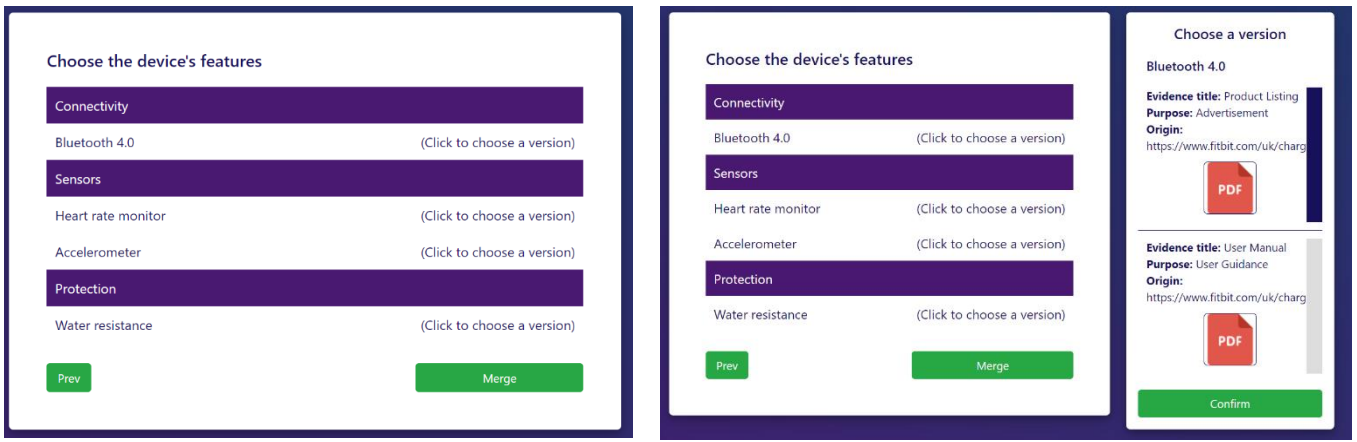

*Figure 5: Learners are asked to merge each feature one by one in order to create a master IoT product profile. This process aims to initiate discussion between team members who are involved in merging: Each team member ideally should have contributed by creating their own IoT product profile in stage 1*

**DESCRIPTION OF EXPERIMENT**

We conducted two different user studies to examine the effectiveness of our proposed learning activity. We selected two different user groups, namely, (1) Youth Group and (2) Professionals. Both of our studies were aimed to gather qualitative information. We used open mailing lists to requires the participants. In this context, learners are called participants.

- **User Study 1: Youth Group:** This study was performed on a group of college students from a youth group. The sample comprised of 6 students, aged between 16-18 years old, all with very limited technical knowledge beyond what mobile phone they carry and what video games console they prefer. Beyond ICT lessons back in school, the students had little experience with computers. All had never heard of IoT by name but were familiar with some mainstream devices, such as the Fitbit. Despite this, none of the students was in possession of such a device or knew anyone that did. These users will be identified from now as YG1-YG6. Our intention of this study was to explore, how can we use a learning activity like ours to increase the awareness (their features and design decision) of IoT products among high school students.

- **User Study 2: Professionals:** This study was performed on a group of 8 participants. They range in age and occupation.

*Table 1: User Study 2 - Participants*

| User ID | Age   | Occupation      |
|---------|-------|-----------------|
| P1      | 30-40 | Teacher         |
| P2      | 18-25 | Junior Doctor   |
| P3      | 18-25 | Nutritionist    |
| P4      | 30-40 | Doctor          |
| P5      | 18-25 | Civil Engineer  |
| P6      | 30-40 | Doctor          |
| P7      | 30-40 | Accountant      |
| P8      | 18-25 | Accountant      |

**Procedure**

We formulated the user studies into four phases, where each is serving its purpose. We audio recorded all the discussions and analysis is presented later in the paper.

**Phase 1 – Mobile Phones:** In the first phase, participants were asked to list and rank features and characteristics they consider when purchasing a mobile phone. This was done to establish the foundation of their technological mental model, so later they can try to the extent it to the IoT domain. All of our participants were familiar with mobile phones. Therefore, this exercise acted as an ice breaker. It also helps to build confidence and to bring some familiarity to the user study. We anticipate that asking questions directly related to IoT would make participants uncomfortable as they may have a clear reference point to start the thinking process.

**Phase 2 – IoT:** In this phase, participants were introduced to IoT. They were asked whether they had any knowledge or experience of IoT (Smart wearables and Smart Homes). Definitions and examples were then explained to them, where they were then asked if they owned such a device, without knowing it belonged to IoT. Next, participants were asked to research one activity tracker – *Fitbit*, *Beddit*, *VivoSport* - then asked to further develop their mental models by listing and ranking the features they believed would be important to them. Participants did not engage with our propose platform yet.

**Phase 3 – Using the OLYMPUS Platform:** In this phase, participants were introduced to the **OLYMPUS** platform and asked to complete the three-stage learning process. **[Stage One – Crowdsourcing]** Participants were split into groups that each participant acted the role of a crowd worker. In each team, participants were asked to individually research, investigate, analyse the IoT product's details and features. **[Stage Two – Merging]** Once all members of a group completed their revisions, as a group, they were asked performed a merge, creating a master profile of the IoT product. **[Stage Three – Comparing]** Learners then used the compare feature (i.e., login as students) of the platform to compare six other devices already entered into the system. During this stage, users were asked to note down and discuss any surprises they had and anything they learnt while comparing.

**DISCUSSIONS AND LESSONS LEARNT**

We followed Miles' framework [11] to conduct qualitative analysis. Further, for data reduction phase, we use Richards' three-tier coding technique (i.e., descriptive coding, topic coding, and analytic coding) [12]. The thematic areas we found by analysing the data across both studies are as follows:

- **Learners struggled to relate when lacking in either knowledge or past experiences**

Having never heard of IoT and never been in a position of purchasing or requiring an activity tracker, the youth group struggled to envisage a situation where they were purchasing one. Youth group struggled to relate to the described situation, having never been in one previously; unlike the exercise in phase 1 – mobile phones – where they were perfectly comfortable listing and ranking their considerations. In addition, the youth group members knew very little about the sort of features within a typical activity tracker, so this coupled with the difficulty in relating, meant they struggled to come up with characteristics they would consider.

Despite this being the first time, the professional had heard of IoT, they were able to grasp the concept very quickly, with two users realising they owned a few devices – an *Amazon Echo* and a *Fitbit*. Interestingly, this suggests that while technology has become intertwined with our lives, many people don't understand how these devices have become so useful – what sort of data is stored and how it processes and manipulates it to *better* our lives.

In addition, these learners were relatively knowledgeable with respect to technology. P1 was able to distinguish the difference between Bluetooth and WIFI, "*Bluetooth is for short distances, whereas WIFI is for longer distances.*" Furthermore, P8 was able to identify four sensors within a mobile phone, "*pressure, temperature, GPS and accelerometer",* as well P2 being able to identify another mobile connectivity medium, NFC – "*used for mobile payments*". This shows the users had a reasonable understanding of the handsets they use hourly. With this knowledge, we believe they found relating to the proposed scenario much easier.

The OLYMPUS platform was developed to support the verity of different audiences. Based on above to learner groups, it safe to say that better the IoT background and experience (e.g., have used some IoT devices despite they know how they work or not) leaners have, better the would engage with the learning activity (and the platform).

- **Learners gained knowledge on IoT products through the learning activity**

As said earlier, learners completed phase 1 very well, indicative of knowledge they possessed on mobile phones. Following this, learners were unable to define IoT, with some struggling to complete phase 2 of the study, lacked the knowledge and experience. However, during the final stage of the learning process, comparing devices, YG1 expressed "*surprise*" when comparing a Google Home Mini with an Amazon Echo, claiming *"[they] thought the Amazon Echo would have WIFI."* The learner's statement shows that they had drawn a comparison between the two devices – proving a level of understanding of the two devices - and recognised WIFI is an essential feature with regards to connectivity, replying, "*so you can connect to your mobile or your lights*" in response to being asked why WIFI would be important to an Amazon Echo or Google Home. Learners, P7 and P8, expressed their surprise "*by the number of sensors in a June Oven"*, expecting it to contain far less.

Following the comparison stage, we began by asking a number of questions relating to mobile phones, so learners could draw parallels when asked similarly for IoT devices. As mentioned previously, learners were able to differentiate between Bluetooth and WIFI, as well as identify NFC. Despite the simplicity of these answers, it showed learners understood the basics – enough information for the average person to decide what technology to use for two scenarios – short distances and longer distances. This knowledge could very well be applied when purchasing IoT devices. For instance, if a device only has Bluetooth, while another has both Bluetooth and WIFI, consumers with this knowledge, could make informed decisions. In another instance, learners, P3 and P8 demonstrated their learning, by answering a question asked by fellow learners, P5, "w*hy does it [June Oven] have a camera",* to which P3 replied with, "*cause you can watch the food on the app, to see how well done it [the food] is*," and P8, "*to recognise food and [determine] how long it should be in for."* Overall discussion, based on the above evidence, seems to have improved the IoT knowledge of our learners.

- **Security and Privacy Concerns**

Another indicator of successful learning is the application of said knowledge. Having started by asking learners what familiar technologies, like Bluetooth and WIFI, are used for, we moved onto some applied thought-provoking questions, such as, "in terms of privacy, rank these devices in order of

danger [Amazon Echo, Beddit, Fitbit, Google Home, June Oven, Oral-B Smart toothbrush]." The consensus was:

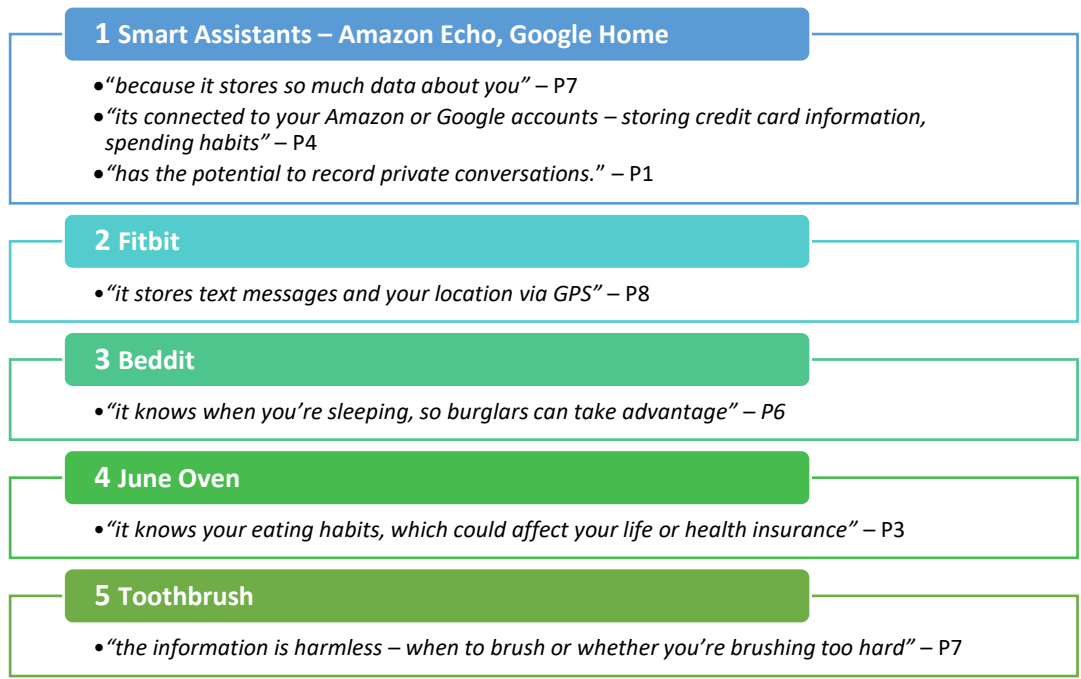

*Figure 6: Ranked Perceived Privacy Risk*

1. **Smart Assistants – Amazon Echo, Google Home**
   - *"because it stores so much data about you"* – P7
   - *"its connected to your Amazon or Google accounts – storing credit card information, spending habits"* – P4
   - *"has the potential to record private conversations."* – P1

2. **Fitbit**
   - *"it stores text messages and your location via GPS"* – P8

3. **Beddit**
   - *"it knows when you're sleeping, so burglars can take advantage"* – P6

4. **June Oven**
   - *"it knows your eating habits, which could affect your life or health insurance"* – P3

5. **Toothbrush**
   - *"the information is harmless – when to brush or whether you're brushing too hard"* – P7

Upon being asked how they came to this ranking, P7 explained, "it's about the quality and sensitivity of the information they store that determines the danger of the device." To further validate their reasoning, learners were asked, whether the rankings would change, if the toothbrush tracked your location, to which, P8 replied, "*yes, as its private information, but still lower down the table, as the others store more forms of private information.*" This statement implies that this learner had gained an insight into the sort of information the devices stored and was able to create a risk assessment internally, should the information be leaked.

To further understand where their priorities lied, we asked the learners to shed some light onto what details do they consider to be most private: "*bank details*", (P4) "*private conversations*" (P8), "*well just about anything personal*" (P7), which begged the question, why would you trust such a device with this sort of information? P3 explained, "*most people are unaware of what these devices can do, what they store and how they use it. I think the usefulness of a device can sometimes outweigh the worry of sharing the information.*" Interestingly, not all companies disclose what information can be derived from a device's sensors. That's both the blessing and the curse of technology, finding new and intuitive ways to mine data, without being too intrusive, but as a consumer, you're unknowingly sharing information you probably didn't want to. P6 described this perfectly, "*we don't know how these sensors work, and come up with some of their statistics or recommendations, we only came up with our rankings based on what we thought the sensors do.*"

**CONCLUSIONS AND FUTURE WORK**

Throughout this paper, we presented a learning method we developed to support IoT education. We formulated our approach around crowdsourcing technique. To facilitate our method, we develop a crowdsourcing platform called OLYMPUS. To validate our approach, we conducted two user studies. These user studies shed some light on how a platform like ours could supplement the IoT learning process. Through the two user studies, we gather numerous evidence that demonstrates how the learners benefited by the proposed technique. In the future, we aim to extend this platform in two different ways. First, crowdsourcing techniques can be used to support more formal IoT education (e.g. used to teach IoT system design in a university setting). In such a scenario, learners will attempt to reverse engineers an actual IoT products (e.g., using data flow diagrams and Unified modelling language diagram techniques). Secondly, we aim to extend to OLYMPUS to capture GDPR related information for each IoT product. For example, we may utilise crowdsourcing techniques to extract GDPR related information out of terms and condition documented of each IoT product. In this way, learners may gain better knowledge of privacy by design in IoT.

**ACKNOWLEDGEMENTS**

We acknowledge the funding received by EPSRC PETRAS 2 (EP/S035362/1) and RiR (EP/T517203/1). The learning platform we developed is open source and available for the community: https://gitlab.com/IOTGarage/iot-product-catalogue

**REFERENCES**

[1] C. Perera, C. H. Liu, and S. Jayawardena, "A Survey on Internet of Things From Industrial Market Perspective," *IEEE Access*, vol. 2, pp. 1660–1679, 2014.

[2] C. Perera, C. H. Liu, and S. Jayawardena, "The Emerging Internet of Things Marketplace from an Industrial Perspective: A Survey," *IEEE Trans. Emerg. Top. Comput.*, vol. 3, no. 4, pp. 585–598, 2015.

[3] P. Kucherbaev, F. Daniel, S. Tranquillini, and M. Marchese, "Crowdsourcing Processes: A Survey of Approaches and Opportunities," *IEEE Internet Comput.*, 2016.

[4] Jeff Howe, "The Rise of Crowdsourcing | WIRED," *WIRED*, 2006. [Online]. Available: https://www.wired.com/2006/06/crowds/. [Accessed: 20-May-2019].

[5] E. A. Alghamdi, N. R. Aljohani, A. N. Alsaleh, W. Bedewi, and M. Basheri, "CrowdyQ: a virtual crowdsourcing platform for question items development in higher education," in *Proceedings of the 17th International Conference on Information Integration and Web-based Applications &Services - iiWAS '15*, 2015, pp. 1–4.

[6] A. Ranj Bar and M. Maheswaran, *Confidentiality and Integrity in Crowdsourcing Systems*. Cham: Springer International Publishing, 2014.

[7] D. C. Brabham, *Crowdsourcing*, The MIT Pr. MIT Press, 2013.

[8] E. Estellés-Arolas and F. González-Ladrón-de-Guevara, "Towards an integrated crowdsourcing definition," *J. Inf. Sci.*, vol. 38, no. 2, pp. 189–200, Apr. 2012.

[9] J. Dunlosky, K. A. Rawson, E. J. Marsh, M. J. Nathan, and D. T. Willingham, "Improving Students' Learning With Effective Learning Techniques," *Psychol. Sci. Public Interes.*, vol. 14, no. 1, pp. 4–58, Jan. 2013.

[10] J. B. (John B. Biggs and K. F. (Kevin F. Collis, *Evaluating the quality of learning : the SOLO taxonomy (structure of the observed learning outcome)*. Academic Press, 1982.

[11] M. B. Miles, A. M. Huberman, and J. Saldaña, *Qualitative data analysis : a methods sourcebook*. 2013.

[12] L. Richards, *Handling qualitative data : a practical guide*. 2014.